\documentclass[prl,showpacs,twocolumn]{revtex4}
\usepackage{mathrsfs}
\usepackage{amsmath}
\usepackage{amssymb}
\usepackage{revsymb}
\usepackage{graphicx}
\usepackage{mathrsfs}

\begin{document}
\title{Zeptosecond $\gamma$-ray pulses} 
\author{Michael Klaiber$^{1,2}$}
\email{klaiber@mpi-hd.mpg.de}
\author{Karen Z. Hatsagortsyan$^1$}
\email{k.hatsagortsyan@mpi-hd.mpg.de}
\author{Christoph H. Keitel$^{1}$}
\email{keitel@mpi-hd.mpg.de}
\affiliation{$^1$Max-Planck Institut f\"ur Kernphysik, Saupfercheckweg 1, D-69117 Heidelberg, Germany\\
$^2$Theoretische Quantendynamik, Physikalisches Institut der Universit\"at, H.-Herder-Str. 3, D-79104 Freiburg, Germany}

\date{\today}

\begin{abstract}

High-order harmonic generation (HHG) in the relativistic regime is employed to obtain zeptosecond pulses of $\gamma$-rays. The harmonics are generated from atomic systems in counterpropagating strong attosecond laser pulse trains of linear polarization. In this setup recollisions of the ionized electrons can be achieved in the highly relativistic regime via a reversal of the commonly deteriorating drift and without instability of the electron dynamics such as in a standing laser wave. As a result, coherent attosecond $\gamma$-rays in the $10$ MeV energy range as well as coherent zeptosecond $\gamma$-ray pulses of MeV photon energy for time-resolved nuclear spectroscopy become feasible.

\end{abstract}

\maketitle

The time-resolved monitoring of fast-evolving processes with the pump-probe technique requires short laser pulses. The dynamics of chemical reactions has been probed by sub-picosecond laser pulses \cite{Zewail}. Strong infrared laser pulses of femtosecond duration have been used to investigate the time evolution of vibrational wave packets in molecules \cite{Ullrich}. The new emerging techniques for generation of attosecond pulses of extreme ultraviolet radiation \cite{atto} and of recolliding electron wavepackets \cite{Niikura} have been exploited to look into the Auge-decay process \cite{Drescher}, the nonlinear ionization dynamics \cite{as}, molecular dynamics \cite{Lein} and in theory to track the motion of an electronic wave packet in an atom \cite{Hu}. The required frequencies of the short pulses depend on the characteristic energies of the processes by means of which the fast dynamics is governed. While chemical reactions can be controlled with a few eV excitations driven by a weak infrared laser field via one-photon processes, the molecular dynamics can be mediated with a few 10 eV ionization transitions by a strong infrared laser field via multiphoton processes, and photon energies from a few 100 eV  up to several keV are required to control the inner-shell electron dynamics.

The time-resolved investigation of nuclear processes is a challenging problem \cite{Ledingham}. How large photon energies and how short photon pulses are required to deal with this task? It is known \cite{nuclear_physics} that typical energies of nuclear single-particle transitions are of order of $1-10$ MeV with typical decay lifetimes of the levels of around $10^{-9}-10^{-15}$ s. The energies of the collective nuclear excitations range from several
dozens of keV up to $30$ MeV; the electromagnetic giant dipole resonance is at about $15-22$ MeV (the width of the resonance is of the order of $2-7$ MeV). This sets the scale for the required photon energies. The disintegration time of the compound nuclei during nuclear reactions ranges  from $10^{-19}-10^{-16}$ s. Some nuclear processes, such as the decay of excited levels, can be tracked with pulses longer than $100$ fs for which the synchrotron radiation sources are well suited. However, there is a wealth of nuclear phenomena for which the investigation of the time resolved  dynamics requires much shorter photon pulses of up to a few zeptoseconds (zs) duration, such as e.g. resonance fluorescence \cite{Arad} ($1$ fs timescale), resonance internal conversion \cite{Karpeshin} ($1$ as timescale), compound nuclei evolution \cite{Maruyama} or photodisintegration of nuclei \cite{Utsunomiya} (zs timescale).

One of the successful ways towards coherent high frequency and ultrashort pulse production is connected with HHG \cite{atto}. The highest photon energy achieved via HHG in gas jets today is at about 2.5 keV \cite{Seres} and the shortest pulse length achieved in the same way is 130 as \cite{Sansone}. To obtain higher photon energies, relativistic laser intensities are required. However, the relativistic drift of the ionized electron suppresses the HHG yield in atomic systems \cite{review}. In the weakly relativistic regime there are methods to counteract the drift \cite{drift} such that higher photon energies but not necessarily short coherent laser pulses are feasible. State-of-the-art proposals with zeptosecond pulses reach keV photon HHG via overdense plasmas in ultrarelativistic laser pulses 
\cite{Tsakiris}.

\begin{figure}
  \begin{center}
    \includegraphics[width=0.35\textwidth,clip=true]{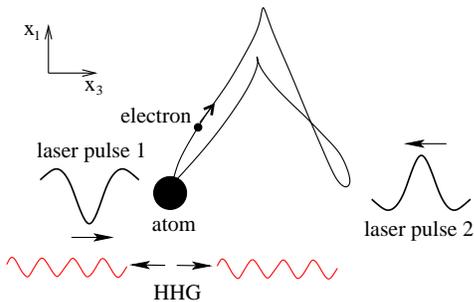}
\caption{With two counterpropagating APTs rather than a conventional
sinusoidal pulse, it is possible to essentially revert the relativistic drift of the ionized electron
and to enable electron rescattering  in the highly relativistic regime (see the electron trajectory).  x and z are the laser polarization
and propagation directions of the APT, respectively. }
    \label{Trajec}
  \end{center}
\end{figure}

In this letter we show  that coherent $\gamma$-rays up to the $10$ MeV energy range as well as zeptosecond pulses of $\gamma$-rays of MeV photon energy are feasible to allow  for time-resolved nuclear spectroscopy. This is  realized via relativistic HHG with highly charged ions in counterpropagating strong attosecond pulse trains (APTs) of linear polarization (see Fig. \ref{Trajec}). In this setup  the common problems with the relativistic drift of the ionized electrons at cutoff energy are circumvented: the electron is ionized by one attosecond pulse, driven by it up to the end of the pulse, then accelerated by the following counterpropagating pulse which reverts the drift and realizes the rescattering with the atomic core.

\begin{figure}
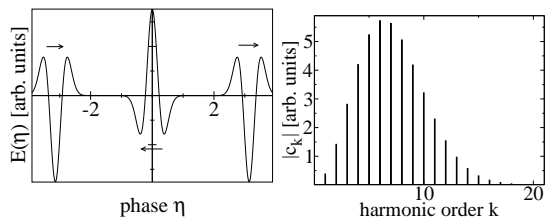

  \begin{center}
\includegraphics[width=3.5cm]{klaiberfig2a.eps}
\includegraphics[width=3.5cm,clip=true]{klaiberfig2b.eps}
\caption{(a) Phase dependence of the electric field ($E(\eta)$) of the counterpropagating APTs. The repetition angular frequency of the pulses in the train is $\omega =0.05$ a.u.. The propagation direction of the pulses is shown by the arrows; 
  (b) Frequency spectrum ($|c(k)|$) of one APT. The frequency components are phase locked.}
\label{pulse}
  \end{center}
\end{figure}

We consider the HHG process of an atomic system driven by counterpropagating APTs which arise most favorably from the same beam via a beam splitter. Our investigation of HHG is based
on the solution of the Klein-Gordon equation in the strong field approximation (SFA) \cite{sfa}. 
The APTs are linearly polarized with vector
potentials
$\mathbf{A}_1(x)=A(\pi+\eta_1)\mathbf{e}_x$
and 
$\mathbf{A}_2(x)=-A(\eta_2)\mathbf{e}_x$, with
$\eta_1=t-z/c$, $\eta_2=t+z/c$ and the 
unit vector in the polarization direction $\mathbf{e}_x$. The amplitude of HHG within the SFA  in the single-active electron approximation is given by the following expression
\cite{rel_hhg} (atomic units are used throughout the paper):
\begin{eqnarray}
  M_{n}&=&
    -i \int
    d^4x^{\prime}\int d^4x^{\prime\prime} \left\{\Phi(x^{\prime})^*
    \right. \nonumber \\
      &\times &   \left. V_{H}(x^{\prime})G(x^{\prime},x^{\prime\prime})
     V_{A}(x^{\prime\prime})\Phi(x^{\prime\prime})\right\}
     \label{mhhg}
\end{eqnarray}
with harmonic order $n$, 
$V_H(x)=2\mathbf{A}_H(x)/c\cdot(\hat{\mathbf{p}}+\mathbf{A}_L(x)/c)$, $V_{A}(x)=2iV(x)/c^2\partial_{t}+V(x)^2/c^2$, atomic potential $V(x)$, momentum
operator of the electron $\hat{\mathbf{p}}$, vector potential
of the overall laser field in the radiation gauge $\mathbf{A}_L=\mathbf{A}_1+\mathbf{A}_2$,  matrix element
of the vector potential of the high harmonic field for an one-photon emission process $\mathbf{A}_H(x)$, bound state wave function $\Phi (x)$ as an eigenstate of the energy operator in the radiation gauge \cite{klaiber1}, time-space coordinate $x=(ct,\mathbf{x})=(ct,x,y,z)$, and speed of light $c$. In Eq.(\ref{mhhg}) $G(x^{\prime},x^{\prime\prime})$ represents 
the Green function of the Klein-Gordon equation for the electron in the field of both APTs. There are two different scenarios for the electron dynamics after ionization. In one scenario the electron successively moves in different counterpropagating pulses. In the other scenario, the two counterpropagating pulses act on the ionized electron simultaneously during its excursion time. The recollisions of the ionized electrons with non-vanishing probability and HHG are connected only with the first scenario of the interaction in the highly relativistic regime because of chaotic electron dynamics and a negligibly small range of electron phases for rescattering in the second scenario \cite{standing_wave}. We thus restricts ourselves to situations where the  first scenario applies.
Without loss of generality the process is initiated by the action of the first laser field followed by the second one.  Then, the Green function of the Klein-Gordon equation for the electron in the field of both APTs can be approximated  via the Klein-Gordon Volkov Green
function for a single laser field $G_i^V(x^{\prime},x^{\prime\prime})$ ($i \in \{1,2\}$ refering to the respective APT) in the following way:
\begin{eqnarray}
  G^V(x,x^{\prime})=i \int d^3\mathbf{x}^B
  G^V_2(x,x^{B})\overset{\leftrightarrow}{\partial}_{ct^B}G^V_1(x^B,x^{\prime})
  \label{mhhg1}
\end{eqnarray}
with  time $t^B$ such that the first laser pulse has already left the wave packet of the active electron and the second laser pulse has not acted yet on the electron at space coordinate $\mathbf{x}^B$. As in the parameter range of interest here $K\omega \ll I_p$ with the largest frequency component of the attosecond pulse $K$, repetition angular frequency in the APT $\omega$ and ionization potential $I_p$, the integral in Eq.(\ref{mhhg}) can be calculated via the saddle-point method.

\begin{figure}
  \begin{center}
   \includegraphics[width=0.35\textwidth,clip=true]{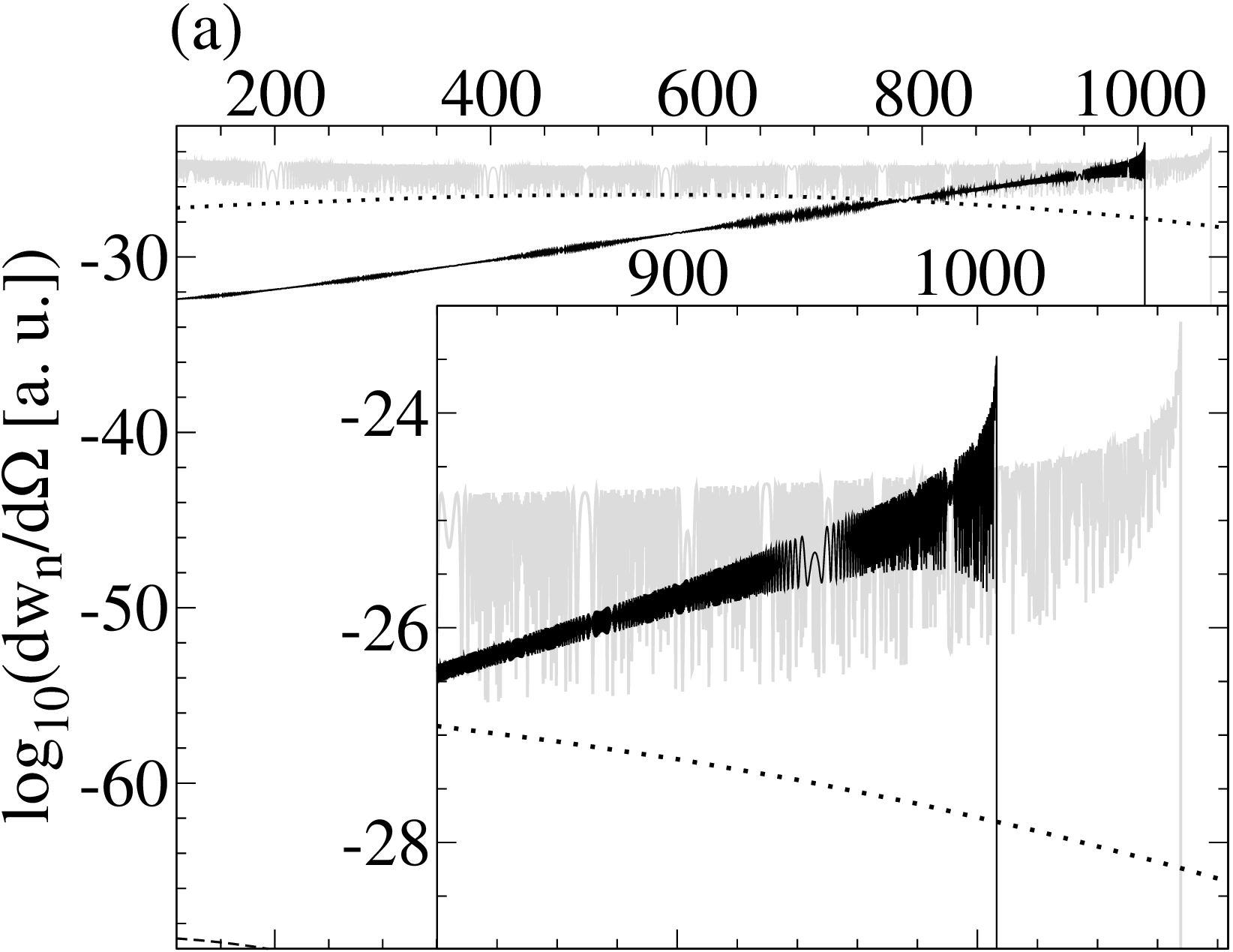}
    \includegraphics[width=0.35\textwidth,clip=true]{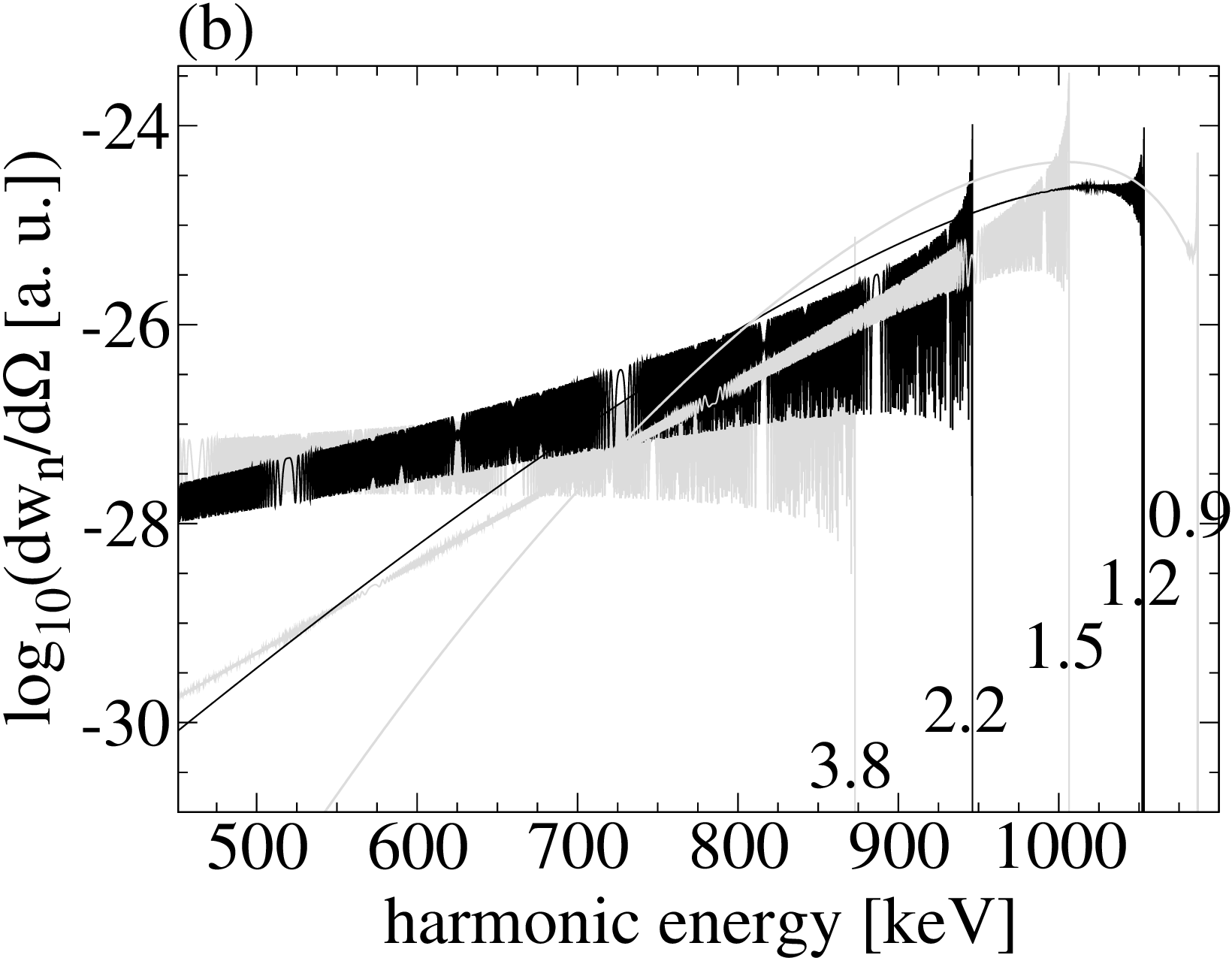}
            \caption{Harmonic emission rate in APT propagation directions 
      as function of the harmonic energy ($dw_n/d\Omega$ is the emission probability of the $n^{th}$ harmonic per unit time and per unit solid angle $\Omega$). The laser field amplitude is $E_0=88$ a.u., $I_p=63$ a.u. (Mg$^{10+}$): \\
(a) (solid black) two counterpropagating APTs via the Klein-Gordon equation, (solid gray) two counterpropagating APTs within the DA, (dotted)  tailored APT as in \cite{klaiber2} via the Klein-Gordon equation, (dashed) two copropagating APTs with the pulses as in Fig. \ref{pulse} via the Klein-Gordon equation (see very bottom left), $\omega=0.05$ a.u.. The window shows enlarged the cutoff energy region. (b) The dependence of HHG yield on the time delay between the pulses; the  delays are indicated for each spectrum in units of fs.}
    \label{h}
  \end{center}
\end{figure}

In Fig. \ref{h} the HHG yield is shown in the setup of counterpropagating APTs for a pulse shape as in Fig. \ref{pulse}. The yield is compared with the spectrum calculated in the dipole approximation (DA) for the setup of two counterpropagating APTs as well as with
the result of the specially tailored APT as in \cite{klaiber2} with pulses of rectangular shape. The DA result can be regarded as a reference for the ionization-rescattering process without drift, since in this description the 
relativistic drift is omitted. In the considered strongly relativistic regime with a laser field amplitude of $E_0=88$ a.u.  (peak intensity $6\times 10^{20}$ W/cm$^2$), 
the counterpropagating APT setup 
produces harmonics with high efficiency, matching the DA result, in the cutoff region of $1$ MeV energy. It shows also a strong improvement with respect to the proposal which employs a single tailored APT \cite{klaiber2}, gaining up to three orders of magnitude in the HHG rate. Moreover, the conditions for the pulse features in the present setup are less demanding as those for the tailored pulses in \cite{klaiber2}. In the setup here each pulse in the train is of about $300$ as duration and contains less than 20 harmonics, while for the single tailored APT up to 100 harmonics are required. Note that the severe requirements for the special pulse tailoring is absolutely necessary in the case of a single APT as the negligible HHG yield in the pulse without tailoring in Fig. \ref{h}(a) shows.

The dependence of the HHG spectra on the pulse separation is presented in Fig.~\ref{h}(b).
The HHG spectra are characterized by a bent plateau
that ends in a sharp cutoff. The bending increases for 
shorter time delays between the pulses, whereas the oscillation pattern due
to the interference of two possible quasiclassical trajectories decreases. Both
features indicate  stronger relativistic signature of the process for shorter time delays \cite{rel_hhg}. 
The longer the time interval between the two pulses, the more excursion space is given to the electron and the recolliding electron need to ionize at higher field strengths resulting in higher ionization and thus higher HHG rate. Additional quantum spreading of the ionized electron can reverse this effect for long time delays as visible from the spectrum. 
Further,  the cutoff energy is slowly increasing with a reduction of the time delay. This is because the ionization of the recolliding electron tends to occur at a later moment in the first pulse if the time delay is shorter. As a result, due to the particular shape of the pulses in Fig. \ref{pulse}, the energy of the electron after the first pulse is larger and the time left for the final acceleration in the second pulse is higher, too.
Both dependences, the increase of the ionization rate as well as the decrease of the cutoff energy become weaker for longer time delays.

\begin{figure}
  \begin{center}
   \includegraphics[width=0.35\textwidth,clip=true]{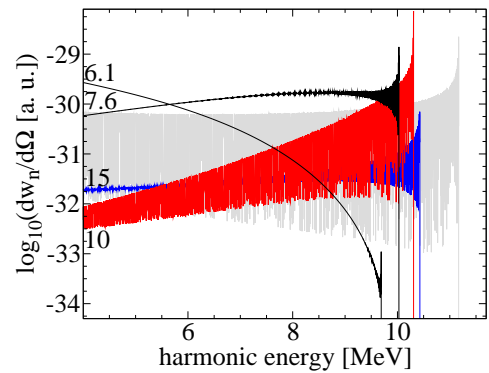}
   \caption{(Color online) Harmonic emission rate in APT propagation directions 
      as function of the harmonic energy for counterpropagating APTs with various time delays 
      compared with the spectrum in the DA (pulse delay $10$ fs) (gray). The pulse delays for the corresponding spectrum are indicated in units of fs; further, $E_0=325$ a.u. and $I_p=150$ a.u. (Ar$^{16+}$).}
    \label{hx325}
  \end{center}
\end{figure}

The counterpropagating APTs setup enables the extension of efficient HHG into the highly relativistic regime.
Thus, in Fig.~\ref{hx325} we show HHG spectra employing APTs with $E_0=325$ a.u. (peak intensity $8\times 10^{21}$ W/cm$^2$) and various time delays between the APTs.
The harmonic emission rate in the cutoff region is as intense as the one in the DA, i.e. the recollision can be realized even in the highly relativistic regime. Then, HHG cutoff energies of about 10 MeV with a significant emission rate are possible. 
The rate decrease compared with the previous case of Fig. \ref{h} is only due to quantum spreading which is enhanced because of the stronger laser electric field and the longer time delay between the pulses. The latter is necessary for reverting the relativistic drift. 
Further, the reduction of the rate at cutoff energy, when decreasing the time delay from $10$ fs, is larger in this regime than for weaker laser fields. The cutoff energy dependence on the time delay is inverted compared to the weak field case due to the stronger relativistic signature of the process. An optimal time delay is at about 8 fs.

\begin{figure}
  \begin{center}
   \includegraphics[width=0.3\textwidth,clip=true]{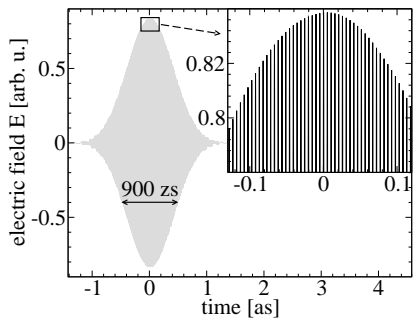}
   \caption{Temporal shape of the generated pulse with a Gaussian window at 805 keV with a FWHM of 4 keV. The FWHM of the pulse is approximately 900 zs. $E_0=88$ au., $I_p=63$ a.u. and a time delay between the pulses of $0.9$ fs. }
   \label{zepto}
 \end{center}
\end{figure}

In the relativistic regime, a broad spectral content of the emitted harmonic radiation as well as the fact that mainly one trajectory contribute to HHG, are beneficial for ultrashort pulse generation. 
An analysis of the harmonic phases shows that the variation of the rescattering phase  by $0.35$ yields a variation of the rescattering energy of $1000$ keV. Then the spectral window to generate the shortest possible pulse is at about a few keV. Choosing a $4$ keV window near the spectral region of  $1$ MeV where the HHG yield is largest, a zeptosecond pulse with FWHM of $900$ zs can be produced (see Fig. \ref{zepto}).
The zeptosecond pulse in the considered case is rather weak: $N_{\gamma}=10^4$ coherent photons per pulse are feasible when an interaction length of $1$ cm, a cross-section of $1$ mm$^2$ and an ion density of $10^{19}$ cm$^{-3}$ (e.g. in the "bubble" regime of laser-plasma interaction \cite{Leemans}) are employed. However, due to the shortness of the pulse, the peak flux density of the photons is rather large: $10^{22}$ photons$\cdot$ s$^{-1}$mm$^{-2}$ \cite{flux}. On roughly estimating the number of interaction events $N=\sigma N_{\gamma}\rho_nL_{int}$ between zeptosecond pulse and nuclei, with cross-section $\sigma$ of photonucleus collective interaction, nuclei density $\rho_n$, and interaction length $L_{int}$, we obtain $N\simeq 10^3$ interaction events per zeptosecond pulse with $\sigma \simeq 10^{-26}$ cm$^2$, $\rho_n\simeq 10^{23}$ cm$^{-3}$, and  $L_{int}\simeq 10^2$ cm. The latter indicates at least theoretical feasibility of nuclear spectroscopy with these pulses. We note that the intensity of the zeptosecond pulse could be increased, e.g. by an additional tailoring of the pulses in the APTs.

In conclusion, we have shown that HHG in the relativistic regime with an atomic system can be realized employing counterpropagating strong APTs. Coherent $\gamma$-rays in the $10$ MeV energy range as well as coherent zeptosecond pulses of MeV photon energy can be generated this way. The main difficulty in experimental realization of this setup is related to the availability of strong APTs. Those pulses combine two extreme properties:  short pulse duration and  high intensity. However, recent advances in the field of strong laser field interaction with overdense plasmas show that strong attosecond pulses are expected to be created in the near future \cite{Tsakiris,Pirozhkov}.

Funding by Deutsche Forschungsgemeinschaft via KE-721-1 is acknowledged.


\begin{thebibliography}{99}
\bibitem{flux}   For comparison, the flux density of the  MeV photon source based on  Compton backscattering, being used in \cite{Utsunomiya}, is $10^{4}$ photons$\cdot$ s$\cdot$ mm$^2$.
\end{thebibliography}
\end{document}